# Engineering Non-Hermitian Quantum Evolution using a Hermitian Bath Environment

Mahmoud A. Selim[1], Max Ehrhardt[2], Yuqiang Ding[3], Qi Zhong[3,6], Armando Perez-Leija[3], Konstantinos G. Makris[4,5], Ramy El-Ganainy[6], Şahin K. Özdemir[6], Matthias Heinrich[2], Alexander Szameit[2], Demetrios N. Christodoulides[1,7*], and Mercedeh Khajavikhan[1,7*]

[1]Ming Hsieh Department of Electrical and Computer Engineering, University of Southern California, CA 90089, USA
[2]Institut für Physik, Universität Rostock, Albert-Einstein-Straße 23, 18059 Rostock, Germany
[3]CREOL, College of Optics and Photonics, University of Central Florida, Orlando, Florida 32816-2700, USA
[4]Department of Physics, University of Crete, 70013 Heraklion, Greece
[5]Institute of Electronic Structure and Laser (IESL), FORTH, 71110 Heraklion, Greece
[6]Department of Electrical and Computer Engineering, Saint Louis University, Saint Louis, MO 63103, USA
[7]Department of Physics and Astronomy, University of Southern California, Los Angeles, CA, 90089, USA

* Correspondence and requests for materials should be addressed to Demetrios N. Christodoulides (demetri@usc.edu) (+1 407 721-3811) and Mercedeh Khajavikhan (khajavik@usc.edu) (+1 310 425-9000)

All authors contact details:

Mahmoud A. Selim: maabdelr@usc.edu

Max Ehrhardt: max.ehrhardt@uni-rostock.de

Yuqiang Ding: Yuqiang.Ding@ucf.edu

Qi Zhong: qxz5149@psu.edu

Armando Perez-Leija: aleija@creol.ucf.edu

Konstantinos G. Makris: makris@physics.uoc.gr

Ramy El-Ganainy: ganainy@mtu.edu

Şahin K. Özdemir: sko9@psu.edu

Matthias Heinrich: matthias.heinrich@uni-rostock.de

Alexander Szameit: alexander.szameit@uni-rostock.de

Demetrios N. Christodoulides: demetri@usc.edu

Mercedeh Khajavikhan: khajavik@usc.edu

**Abstract**

Engineering quantum bath networks through non-Hermitian subsystem Hamiltonians has recently emerged as a promising strategy for qubit cooling, state stabilization, and fault-tolerant quantum computation. However, scaling these systems while maintaining precise control over their complex interconnections—especially in the optical domain—poses significant challenges in both theoretical modeling and physical implementation. In this work, drawing on principles from quantum and mathematical physics, we introduce a systematic framework for constructing non-Hermitian subsystems within entirely Hermitian photonic platforms. In particular, controlled exponential decay without actual absorption loss is realized in finite 1-D waveguide chains through discrete-to-continuum coupling and Lanczos transformations. Using this new methodology, we implement parity-time symmetric quantum systems and experimentally demonstrate that these artificial bath environments accurately replicate the dynamics of non-Hermitian arrangements in both single- and multi-photon excitation regimes. Since the non-Hermitian subsystem response deterministically arises from an artificially built Hermitian bath, the quantum evolution can be monitored via post-selection in this fully conservative configuration. This approach bridges the gap between theoretical models and experimental realizations, thus paving the way for exploiting quantum bath engineering in advanced information processing and emerging quantum technologies.

**Introduction**

Historically, efforts in quantum physics have focused on shielding systems from environmental interference, ensuring that coherence is preserved, and the fragile quantum superposition and entanglement remain intact—an essential requirement for quantum computing and precision measurements [1]. In contrast to these strategies, bath engineering has recently emerged as a transformative paradigm for controlling open quantum systems [2,3]. This approach leverages the interplay between coherent quantum dynamics and deliberately tailored dissipative processes to stabilize and manipulate quantum states. Within this framework, while the entire system is still governed by Hermitian principles as dictated by the fundamental tenets of quantum mechanics, the primary quantum subsystem is effectively characterized by non-Hermitian dynamics. Quantum bath engineering has demonstrated efficacy across a broad range of applications, including the stabilization and filtering of entanglement [4,5], and rapid qubit initialization [6]. Similarly, in circuit quantum electrodynamics (cQED), reservoir engineering has proven to be a powerful tool for driving targeted state transitions and facilitating error-resilient quantum operations [7]. In trapped-ion arrangements, this same methodology has been employed to selectively cool specific states and generate high-fidelity entanglement [8], a critical requirement for advancing quantum simulations and computations.

A notable class of open quantum systems is that exhibiting parity–time (PT) symmetry [9,10]. In the classical domain, PT symmetry has facilitated the observation of several counterintuitive phenomena [11–14], including unidirectional invisibility [15], single-mode lasing [16,17], enhanced sensing [18–20], chiral state transfer [21], topological chirality [21–24], and neural networks [25], to name a few. At the core of these advancements is the concept of exceptional points—an intriguing degeneracy in non-Hermitian systems where both the eigenvalues and their corresponding eigenvectors coalesce. In classical optical systems, PT symmetry can be realized by deliberately engineering the refractive index profile such that its real component is spatially symmetric while its imaginary part is antisymmetric [11], i.e., $n_R(x) = n_R(-x)$ and $n_I(x) = -n_I(-x)$. Physically, such implementations can be readily achieved through refractive index modulation and the introduction of optical gain and loss. Nevertheless, extending PT-symmetric dynamics to the quantum realm presents significant challenges, as gain not only introduces noise but can also be further complicated by spontaneous emission. Meanwhile, loss in quantum optics corresponds to photon annihilation, leading to irreversible decoherence. These issues raise fundamental questions as to how non-Hermitian and PT-symmetric dynamics can be utilized in quantum optics. In view of this, one may ask whether these processes can be effectively harnessed by appropriately engineering dissipation. Moreover, is it possible to emulate absorption dynamics in a lossless, all-photonic environment, and can such 'artificial' platforms achieve or surpass the functionalities of their genuine dissipative counterparts under multi-photon excitation conditions? Addressing these questions could pave the way for transformative approaches in the deployment of non-Hermitian strategies within quantum photonic technologies.

In this work, we present a new methodology for realizing a broad class of non-Hermitian quantum Hamiltonians within a fully Hermitian network, drawing on principles from quantum optics and mathematical physics. While this scheme is general, we here specifically focus on parity-time (PT) symmetric Hamiltonians. Inspired by the Wigner-Weisskopf formulation [26], we introduce a systematic approach to engineering the dissipation of a quantum subsystem, even though the photonic network in its entirety is fully Hermitian. In this regard, the conservative nature of this setup allows for precise monitoring of photon dynamics in these Hong-Ou-Mandel multi-port interferometric arrangements [27,28], using deterministic unitary evolution equations, not only in the non-Hermitian subsystem, but also in the bath. We demonstrate the practical implementation of these configurations on an integrated photonic platform and showcase their potential for investigating a range of quantum dynamical phenomena through experimental studies under single- and two-photon excitation conditions.

In 1930, the theoretical work of Wigner and Weisskopf provided a foundational explanation of how states exponentially decay in quantum systems [26]. Their analysis was carried out by considering the interaction between a discrete quantum state, such as an excited atomic level, and a continuum of discrete states. They demonstrated that this interaction leads to a constant decay, which is associated with an exponential decrease over time in the probability of the system remaining in the excited state. The decay rate $\gamma = \pi\kappa^2/\hbar\delta$ is determined by $\kappa$, the constant coupling strength between the excited state and the continuum, while $\delta$ denotes the uniform spacing of the corresponding energy levels [29] (Fig. 1a). This work laid the theoretical foundations for understanding various other decay processes in quantum mechanics, including nuclear gamma decay, autoionization, and the Auger effect. Given the isomorphism between quantum wave mechanics and optical wave dynamics, in principle, this same process can be realized in an all-optical environment by allowing a single waveguide element to interact with an infinite number of mutually uncoupled waveguides through evanescent coupling $\kappa$, as long as their propagation constant is pairwise detuned by an integer multiple of $\delta$, as illustrated in Fig. 1a. As in the Wigner and Weisskopf formalism, if indeed in this scenario the coupling $\kappa$ is constant, one should then anticipate an exponential photon decay in this primary single waveguide element S (Fig. 1a). Unfortunately, while this proposition may seem at first sight conceptually straightforward, it is physically impossible to implement. Obviously, it is impractical to surround a single element with an infinite number of waveguides with variable propagation constants. To circumvent this fundamental obstacle, we systematically deploy a supersymmetric Lanczos algorithm, which reduces a highly interconnected Hermitian Hamiltonian to a tridiagonal matrix that can be readily implemented on a simple one-dimensional chain [30–33] (Fig. 1b). This iterative numerical method serves as a powerful tool for simplifying complex problems that, topologically, may belong to higher dimensions, into a more manageable form.

By employing this approach, one can effectively replicate exponential photon decay dynamics in a prescribed waveguide by adjusting the nearest-neighbor coupling coefficients among the 1D array (bath) elements and their onsite propagation constants, a configuration feasible on several optical platform. In other words, the infinite 2D network shown in Fig. 1a that is responsible for exponential decay can be effectively mapped on an appropriately designed 1D lattice that involves a sufficiently large yet finite number of elements.

## Results

To further elucidate the Lanczos algorithm, let us consider an $m \times m$ Hamiltonian matrix $\mathcal{H}$. For a specific anchor site, the algorithm reduces computational complexity by constructing a reduced tridiagonal matrix $\mathcal{H}_L$ that faithfully captures the essential spectral properties of $\mathcal{H}$ and can be implemented using a one-dimensional array with only nearest-neighbor couplings—an arrangement easily achievable in optical systems via evanescent coupling. The procedure begins by selecting an initial unit vector $|v_0\rangle$, in the $m$-dimensional Hilbert space, to represent a designated anchor site in a multidimensional lattice, and applying $\mathcal{H}$ to obtain the residual vector $|l_1'\rangle = \mathcal{H}|v_0\rangle$. By defining $\epsilon_0 = \langle l_1'|v_0\rangle$, the residual is updated as $|l_1\rangle = |l_1'\rangle - \epsilon_0|v_0\rangle$ to ensure orthogonality. For each subsequent iteration $k = 2, 3, \ldots, m:$, the coefficient $J_k$ is computed as the norm $||l_k||$. The next basis vector is given by $|v_k\rangle = |l_{k-1}\rangle/J_{k-1}$, whereas if $J_{k-1} = 0$, an alternative vector orthogonal to all preceding vectors is chosen. The Hamiltonian is then applied to $|v_k\rangle$ to yield $|l_k'\rangle = H|v_k\rangle$, and the eigenvalue is determined through $\epsilon_k = \langle l_k'|v_j\rangle$. The residual is subsequently updated via $|l_k\rangle = |l_k'\rangle - \epsilon_k|v_k\rangle - J_{k-1}|v_{k-1}\rangle$ to maintain orthogonality. After $m$ iterations, the orthonormal set $\{|v_0\rangle, |v_1\rangle, \ldots, |v_{m-1}\rangle\}$ is assembled into the matrix $V$ whose columns are formed by the vectors $[|v_0\rangle, |v_1\rangle, \ldots, |v_m\rangle]$. In this regard, the corresponding tridiagonal Hamiltonian matrix $\mathcal{H}$ is formed according to:

$$\mathcal{H} = \begin{bmatrix} \epsilon_0 & J_0 & 0 & 0 & \ldots & 0 \\ J_0 & \epsilon_1 & J_1 & 0 & \ldots & 0 \\ 0 & J_1 & \epsilon_2 & J_2 & \ldots & 0 \\ \vdots & \vdots & \vdots & \vdots & \ddots & \vdots \\ 0 & 0 & 0 & 0 & \ldots & \epsilon_{m-1} \end{bmatrix}. \tag{1}$$

where $J_n$ and $\epsilon_n$ refer to the corresponding coupling coefficients and detunings. This tridiagonal matrix preserves the core spectral characteristics of $\mathcal{H}$ while being amenable to experimental realization in a one-dimensional waveguide array setting, thereby significantly reducing fabrication complexity and facilitating the observation of exponential decay dynamics via nearest-neighbor coupling schemes.

The system derived from the Lanczos transformation is intrinsically Hermitian. Specifically, the reduced Hamiltonian $\mathcal{H}$ leads to the quantum evolution equation dynamics for the local bosonic creation operators $a_n^\dagger$ [28,34]:

$$ida_n^\dagger/dz = \epsilon_n a_n^\dagger + J_{n-1} a_{n-1}^\dagger + J_n a_{n+1}^\dagger \tag{2}$$

Through post-selection of outcomes confined to a designated subsystem (here, the first waveguide), Eq. (2) effectively emulates non-Hermitian dynamics. This approach enables the investigation of a wide range of non-Hermitian phenomena on fully Hermitian photonic platforms without introducing external losses that would lead to complete photon annihilation. Furthermore, Eq. (2) can be expressed as $ida_n^\dagger/dz = \mathcal{H} a_n^\dagger$ leading to $a_n^\dagger(z) = e^{-i\mathcal{H}z} a_n^\dagger(0) = U(z) a_n^\dagger(0)$. Here the unitary operator $U(z)$ can be represented as $N \times N$, i.e.,

$$U(z) = \begin{bmatrix} u_{11} & u_{12} & \ldots & u_{1N} \\ u_{21} & u_{22} & \ldots & u_{2N} \\ \ldots & \ldots & \ldots & \ldots \\ u_{N1} & u_{N2} & \ldots & u_{NN} \end{bmatrix} \tag{3}$$

and stands for the transfer matrix of the given system from which one can monitor the quantum evolution via $a_n^\dagger(z) = \sum_i u_{ni} a_i^\dagger(0)$. This formalism is generally applicable under both single- and multi-photon excitations. In other words, the quantum dynamics within this entirely Hermitian framework—encompassing both the subsystem and its engineered environment—can, in principle, be modeled by treating the waveguide array configuration as a multi-port Hong-Ou-Mandel setup. Interestingly, as demonstrated in the following sections, this methodology yields results that are completely aligned with those obtained from the Lindblad master equation:

$$\partial_z \hat{\rho}(z) = -i(\mathcal{H}_{eff}\, \hat{\rho} - \hat{\rho} \mathcal{H}_{eff}^\dagger) + 2\gamma \sum_i \hat{a}_i \hat{\rho} \hat{a}_i^\dagger = \mathcal{L}\hat{\rho}. \tag{4}$$

where $\mathcal{H}_{eff}$ denotes the effective Hamiltonian, $\gamma$ represents the exponential decay factor, and $\hat{\rho}$ represents the reduced density matrix for the PT subsystem. Note that the sum in Eq. (4) only applies for sites displaying a loss factor $\gamma$. In principle, PT symmetric dynamics can be observed in systems solely based on differential loss [35], as for example in a dimer with an effective Hamiltonian $\mathcal{H}_{eff} = \begin{bmatrix} 0 & J_0 \\ J_0 & -i\gamma \end{bmatrix}$. As noted above, a key advantage of employing a fully Hermitian formalism, as in Eq. (2), is that it enables the study and modeling of non-Hermitian dynamics in a universal manner within the Heisenberg picture, without being restricted to the Lindblad framework [36–39]. This is particularly useful in more complicated networks (not displaying exponential decay) where the terms in the Lindbladian can become intractable and not amenable to analysis.

## Materials and methods

To elucidate our methodology, we consider a Wigner–Weiskopf system with a coupling constant $\kappa = 0.11$ cm$^{-1}$ and a detuning parameter $\delta = 0.16$ cm$^{-1}$, which together yield an exponential decay factor $\gamma = 0.25$ cm$^{-1}$. By applying the Lanczos algorithm, we transform the star-shaped configuration into a one-dimensional waveguide array with variable coupling coefficients $J_i$ while intentionally keeping the on-site potentials $\epsilon_i$ the same, as shown in Fig. 2a. These coefficients are obtained from a Lanczos transformation when the system involves a large number of sites $N$ (e.g. $N = 1000$). Note that light leaking into the bath travels a distance of $L_{max} \approx N\pi/(2J_{max})$ before returning back to the subsystem—where $J_{max}$ is the maximum coupling coefficient in the array. In this respect, the artificial environment can then be truncated while preserving the system's response, provided that the system's length is below $L_{max}$. As an example, for the array considered above where $J_{max} = 2.3$ cm$^{-1}$, using $N = 50$ will be sufficient to observe its anticipated behavior up to a propagation length of $z = 25$ cm, as depicted in Fig. 2b.

To experimentally validate our approach, we have implemented an array of 52 evanescently coupled single-mode waveguides using a femtosecond direct writing technique [40]. Ultrashort pulses generated by a frequency-doubled fiber amplifier operating at 517 nm—with a pulse duration of 270 fs and a repetition rate of 333 kHz—are focused inside a 100 mm-long fused silica sample, while the wafer is precisely translated relative to the focal point using a positioning system. This process yields single-mode waveguides with mode field diameters of approximately $13 \times 15$ μm$^2$ with a propagation loss that is below 0.3 dB/cm at the probe wavelength of 810 nm. By varying the separation between adjacent waveguide elements, we tailor the nearest-neighbor coupling coefficients to range from 0.1 cm$^{-1}$ to 2.3 cm$^{-1}$. Furthermore, to facilitate coupling with fiber arrays, fan-out sections are integrated before and after the functional regions of the waveguide array, expanding the channel separations to 127 μm where necessary.

We first characterize the effective loss of a single waveguide attached to an array bath, designed to exhibit an exponential decay constant of $\gamma = 0.25$ cm$^{-1}$, utilizing the setup shown in Fig. 3a. The parameters of the Lanczos waveguide array are identical to those employed in the simulation example described earlier. To observe light evolution, we measured several nominally identical samples of varying lengths using a modified variable stripe technique. Structures ranging from 3 cm to 10 cm in length, in 1 cm increments, were analyzed. Although the overall system remains Hermitian, light propagating in the designated edge waveguide— attached to the artificial environment—displays an exponential decay response, with a relaxation constant $\gamma = 0.25$ cm$^{-1}$ (Fig. 3b). It should be noted that for lengths below 3 cm, Zeno dynamics within the tight-binding model [41] do not yield a purely exponential decay of quantum states. For example, at $z = 0$, the rate of change at the anchor site, $da_1/dz|_{z=0}$, must be zero (see Eq. (2)), whereas the derivative of the exponential function, $de^{-\gamma z}/dz|_{z=0} \neq 0$, does not vanish.

Next, we construct a parity-time dimer system based on the coupling of a neutral waveguide to a "lossy"

channel. The coupling coefficient between the two subsystem waveguide elements is chosen to be $J_0 = 0.3$ cm$^{-1}$, while the attenuation in the lossy channel established through the attached bath is $\gamma = 0.25$ cm$^{-1}$. In this case, the normalized z-dynamics of the PT-symmetric subsystem exhibit a periodic behavior under single photon excitation (Figs. 3c and 3d), consistent with the evolution response before the onset of spontaneous symmetry breaking. Unlike lossless coupled structures, where the rate of power transfer between the two waveguides is identical, here, the transfer rate depends on whether the neutral or the lossy waveguide is initially excited. Specifically, for a Hermitian two-waveguide coupler with $J_0 = 0.3$ cm$^{-1}$, the power transfer occurs over a characteristic length of $L = \pi/2J_0 \approx 5.2$ cm (Fig. 3e), while in the PT-symmetric configuration, exciting the neutral or the lossy waveguide alters this transfer length to approximately 7.3 cm and 4.1 cm, respectively. Figures 3c and 3d shows experimental results when this PT-symmetric system is examined when a single photon is injected in the neutral or the lossy site.

Of particular interest is the response of this artificial PT-symmetric configuration when two-photon states are launched. In our study, entangled photon pairs are generated via spontaneous parametric down-conversion (SPDC) in a periodically poled potassium titanyl phosphate (PPKTP) crystal, with a 405 nm pump photon converted into polarization-entangled signal and idler photons at 810 nm. These photons are separated using a fiber-coupled polarization beam splitter, yielding identically polarized time-frequency correlated photons. Their temporal characteristics are precisely controlled through a Hong–Ou–Mandel (HOM) interferometer [27]. For injecting the $|20\rangle$ and $|02\rangle$ states into PT subsystem, one arm of the HOM arrangement is used as an input, while the other is blocked. The outputs of the waveguides are then connected to single-photon detectors and interfaced with a time tagger for time-correlated single-photon counting. To characterize the evolution of a two-photon quantum state in this artificial PT-symmetric configuration, we again vary the lengths of the waveguide system in the range from 3 cm to 10 cm in 1 cm increments. These results are shown in Fig. 4b. The quantum two-photon excitation response closely matches the theoretical simulations corresponding to a PT coupler, confirming the applicability and versatility of the Lanczos methodology. Notably, even for an initially excited state that is separable (like $|20\rangle$ used here), the quantum evolution can generate entangled states. For a PT system, the maximum degree of entanglement (measured through entropy of the reduced density matrix) occurs when the probabilities satisfy $P_{|11\rangle} = 0.5$ and $P_{|20\rangle} = P_{|02\rangle} = 0.25$. The rate of entanglement generation depends on whether the neutral or lossy waveguide is excited, in stark difference from Hermitian dynamics. For example, for this same system, entanglement is strongly manifested at 3.6 cm and 2.1 cm when exciting the neutral and lossy waveguides, respectively, compared to 2.6 cm in a Hermitian waveguide coupler. This variation highlights the effect of non-Hermiticity on quantum state evolution and entanglement dynamics, which may either accelerate or delay entanglement generation.

In conclusion, in this work we have presented a novel approach for realizing non-Hermitian quantum systems using fully Hermitian photonic structures. This is achieved by means of the Wigner-Weisskopf approach in conjunction with the Lanczos algorithm. This methodology has been successfully used to experimentally characterize the quantum evolution of bi-photon states in an artificial PT symmetric arrangement. The ability to replicate non-Hermitian dynamics in either Markovian or non-Markovian settings, our framework brings a new toolset that can pave the way for advancements in quantum information processing and sensing.

**Acknowledgements**

This work was supported by the Air Force Office of Scientific Research (AFOSR) Multidisciplinary University Research Initiative (MURI) award on Programmable systems with non-Hermitian quantum dynamics (award no. FA9550-21-1-0202) (M.A.S., Y.D., H.M.D., Q.Z., A.P.L., R.E., S.K.O., D.N.C., and M.K.), AFOSR MURI on Novel light-matter interactions in topologically non-trivial Weyl semimetal structures and systems (award no. FA9550-20-1-0322)(M.A.S., H.M.D., D.N.C., and M.K.), ONR MURI award on the classical entanglement of light (award no. N00014-20-1-2789) (M.A.S., H.M.D., D.N.C., and M.K.), The Department of Energy (DE-SC0025224) (M.K., D.N.C., H.M.D. and M.A.S.), W.M. Keck Foundation (D.N.C.), MPS Simons collaboration (Simons grant no. 733682) (D.N.C.), US Air Force Research Laboratory (FA86511820019) (D.N.C.), and the Department of Energy (DESC0022282) (D.N.C. and M.A.S.). The authors thank C. Otto for preparing the high-quality fused silica samples used for the inscription of all photonic structures employed in this work. A.S. acknowledges funding from the Deutsche Forschungsgemeinschaft (grants SZ 276/9-2, SZ 276/19-1, SZ 276/20-1, SZ 276/21-1, SZ 276/27-1, and GRK 2676/1-2023 'Imaging of Quantum Systems', project no. 437567992). A.S. also acknowledges funding from the Krupp von Bohlen and Halbach Foundation as well as from the FET Open Grant EPIQUS (grant no. 899368) within the framework of the European H2020 programme for Excellent Science. A.S. and M.H. acknowledge funding from the Deutsche Forschungsgemeinschaft via SFB 1477 'Light–Matter Interactions at Interfaces' (project no. 441234705). The authors gratefully acknowledge Prof. Yogesh Joglekar for valuable technical discussions.

**Competing interests:** The authors declare no competing interests.

**Author Contributions**
The idea was conceived by M.A.S., A.P.L., D.N.C., and M.K., whereas M.A.S., M.E., A.P.L., Q.Z. Y.D., M.H. designed the structures and fabricated the samples, M.A.S. and M.E. performed the experiments. All authors discussed the results and co-wrote the manuscript.

**References:**

1. C. H. Bennett and D. P. Divincenzo, *Nature* **404**, 247 (2000).

2. J. M. Kitzman, et. al., *Nat. Commun.* **14**, 3910 (2023).

3. A. Shabani and H. Neven, *Phys. Rev. A* **94**, 052301 (2016).

4. M. E. Kimchi-Schwartz, et. al., *Phys. Rev. Lett.* **116**, 240503 (2016).

5. M. A. Selim, et. al., Science **387**, 1424 (2025).

6. J. Tuorila, M. Partanen, T. Ala-Nissila, and M. Möttönen, *npj Quantum Information* **3**, 1 (2017).

7. G. Aiello, M. Féchant, A. Morvan, J. Basset, M. Aprili, J. Gabelli, and J. Estève, *Nat. Commun.* **13**, 7146 (2022).

8. S. Heusler, W. Dür, M. S. Ubben, al -, W. S. Teixeira, M. K. Keller, and F. L. Semião, *New J. Phys.* **24**, 023027 (2022).

9. C. Gaikwad, D. Kowsari, W. Chen, and K. W. Murch, *Phys. Rev. Res.* **5**, L042024 (2023).

10. Y. Ashida, S. Furukawa, and M. Ueda, Nat. Commun. **8**, 15791 (2017).

11. R. El-Ganainy, et. al., *Nat. Phys.* **14**, 11 (2018).

12. C. E. Rüter, K. G. Makris, R. El-Ganainy, D. N. Christodoulides, M. Segev, and D. Kip, *Nat. Phys.* **6**, 192 (2010).

13. K. G. Makris, R. El-Ganainy, D. N. Christodoulides, and Z. H. Musslimani, *Phys. Rev. Lett.* 100, 103904 (2008).

14. K. Özdemir, S. Rotter, F. Nori, and L. Yang, *Nat. Mat.* **18**, 783 (2019).

15. Z. Lin, H. Ramezani, T. Eichelkraut, T. Kottos, H. Cao, and D. N. Christodoulides, *Phys. Rev. Lett.* **106**, 213901 (2011).

16. H. Hodaei, M. A. Miri, M. Heinrich, D. N. Christodoulides, and M. Khajavikhan, *Science* **346**, 975 (2014).

17. L. Feng, Z. J. Wong, R. M. Ma, Y. Wang, and X. Zhang, *Science* **346**, 972 (2014).

18. H. Hodaei, et. al., *Nature* **548**, 187 (2017).

19. Y. H. Lai, Y. K. Lu, M. G. Suh, Z. Yuan, and K. Vahala, *Nature* **576**, 65 (2019).

20. R. Kononchuk, J. Cai, F. Ellis, R. Thevamaran, and T. Kottos, *Nature* **607**, 697 (2022).

21. H. Nasari, et. al., *Nature* **605**, 256 (2022).

22. J. Doppler, et. al., *Nature* **537**, 76 (2016).

23. A. Schumer, et. al., *Science* **375**, 884 (2022).

24. H. Xu, D. Mason, L. Jiang, and J. G. E. Harris, *Nature* **537**, 80 (2016).

25. H. Deng and M. Khajavikhan, *Optica* **8**, 10, 1328 (2021).

26. V. F. Weisskopf and E. P. Wigner, *Z. Phys.* **63**, 54 (1930).

27. C. K. Hong, Z. Y. Ou, and L. Mandel, *Phys. Rev. Lett.* **59**, 2044 (1987).

28. Y. Bromberg, Y. Lahini, R. Morandotti, and Y. Silberberg, *Phys. Rev. Lett.* 102, 253904 (2009).

29. G. Grynberg, A. Aspect, C. Fabre, and C. Cohen-Tannoudji, Introduction to Quantum Optics (2010).

30. B. Cornelius Lanczos, *Journal of research of the National Bureau of Standards* (1950).

31. L. J. Maczewsky, et. al., Nat. Phot. **14**, 76 (2020).

32. J. Huh, S. Mostame, T. Fujita, M. H. Yung, and A. Aspuru-Guzik, *New J. Phys.* **16**, 123008 (2014).

33. U. Weiss, Quantum Dissipative Systems, Fourth Edition 1 (2012).

34. Y. Lahini, Y. Bromberg, D. N. Christodoulides, and Y. Silberberg, *Phys. Rev. Lett.* **105**, 163905 (2010).

35. A. Guo, et. al., *Phys. Rev. Lett.* **103**, 093902 (2009).

36. F. Klauck, L. Teuber, M. Ornigotti, M. Heinrich, S. Scheel, and A. Szameit, *Nat. Photon.* **13**, 883 (2019).

37. W. Chen, M. Abbasi, Y. N. Joglekar, and K. W. Murch, *Phys. Rev. Lett.* **127**, 140504 (2021).

38. I. I. Arkhipov, A. Miranowicz, F. Minganti, and F. Nori, *Phys. Rev. A* **102**, 033715 (2020).

39. H. Moya-Cessa, *Phys. Rep.* **432**, 1 (2006).

40. A. Szameit and S. Nolte, *Journal of Physics B: Atomic, Molecular and Optical Physics* **43**, 163001 (2010).

41. W. M. Itano, D. J. Heinzen, J. J. Bollinger, and D. J. Wineland, *Phys. Rev. A* **41**, 2295 (1990).

### Figure legends

Figure 1 **Lanczos transformation.** The manner a single qubit is coupled to a continuum of discrete levels, as in the Wigner–Weisskopf model **a**, is mathematically isomorphic to that taking place in a star-shape arrangement **b**. **c** Using a Lanczos transformation, the Hamiltonian of these complex systems can be mapped onto a one-dimensional waveguide array with engineered coupling coefficients.

Figure 2 **A Lanczos-designed waveguide array and impact of truncation on PT-symmetric dynamics. a** Coupling coefficients $J_k$ obtained using the Lanczos transformation, aimed to replicate exponential loss at site B via its coupling to the array bath. **b** Evolution of the intensity $I_A$ (site A) in a PT-symmetric coupler for varying number of elements $N$. The effect of truncation for a small number of elements such as $N = 10$ (blue) is obvious. Meanwhile, for a larger number $N = 50$ (purple) the system's response approaches that of the semi-infinite array when the propagation distance is kept below $25\ cm$. Inset: schematic of the PT symmetric system where loss is introduced through the Lanczos array bath.

Figure 3 **Experimental setup and light dynamics in a PT-symmetric Lanczos array. a** Schematic of the experimental arrangement used to observe optical dynamics in an artificial PT

symmetric dimer configuration. This involves a laser diode, a PT symmetric sample, a 4f imaging system, and a detection camera. **b** Exponential decay is observed in a single site A when attached to the Lanczos bath, in full accord with the Wigner–Weisskopf model. **c** and **d** Experimental (markers) and simulated (solid lines) light intensities when sites A and B are separately excited, respectively. In this PT symmetric arrangement, loss is manifested in site B and the coupling between sites A and B is $J_{PT} = 0.3\ cm^{-1}$.

Figure 4 **Two-photon dynamics in a PT-symmetric optical system. a** Experimental setup used. Photon pairs are generated via type-II spontaneous parametric down-conversion (SPDC) using a 405 nm continuous-wave (CW) pump laser and a PPKTP crystal. After spectral filtering and polarization alignment using a fiber polarizing beam splitter (f-PBS), the photons are routed through a delay line into a fiber coupler (FC). One output port is terminated, while the other feeds the photons into the PT symmetric sample. Output light is then collected using a multimode fiber array and detected by single-photon avalanche diodes (SPADs). **b** Measured and simulated dynamics of two-photon states in the PT-symmetric dimer system involving an artificial Lanczos array bath. Left: quantum evolution for an initial state $|20\rangle$ (excitation in the neutral waveguide). Right: similarly for a $|02\rangle$ excitation in the lossy waveguide.

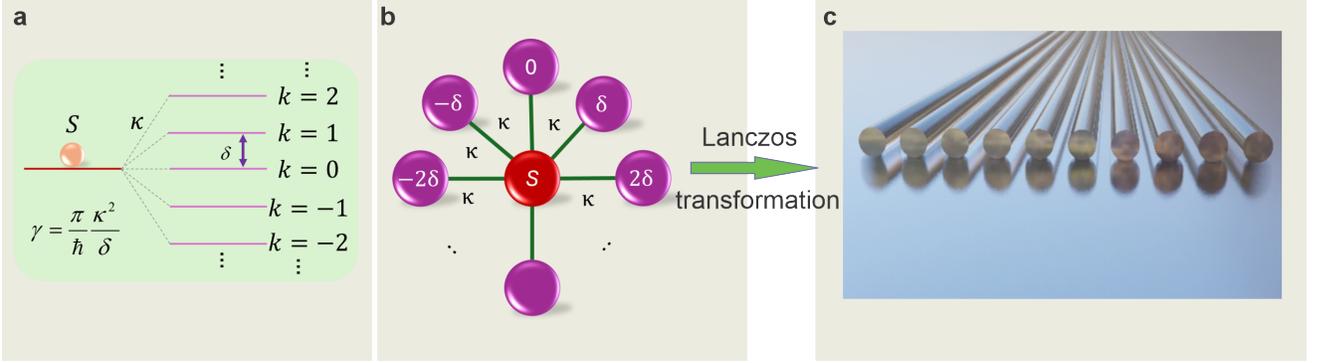

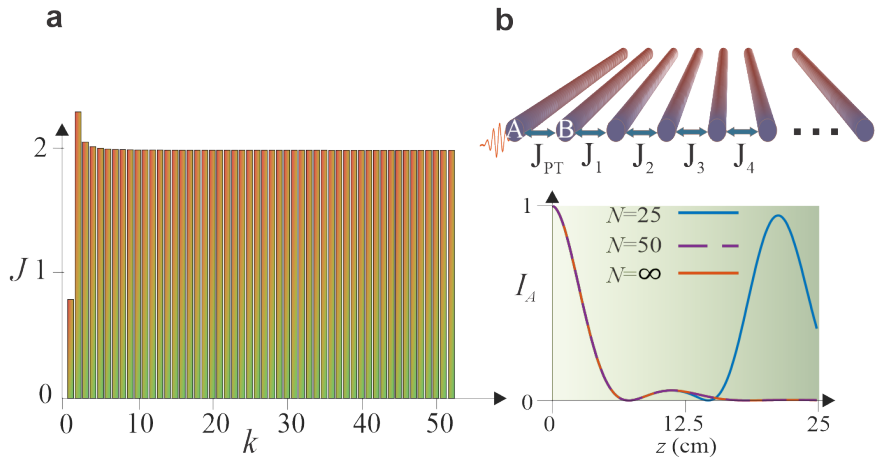

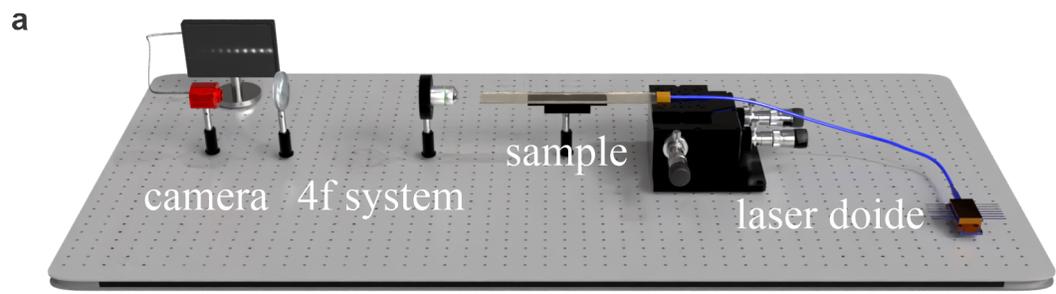

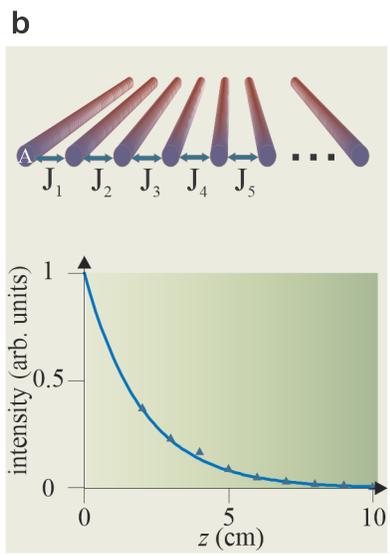

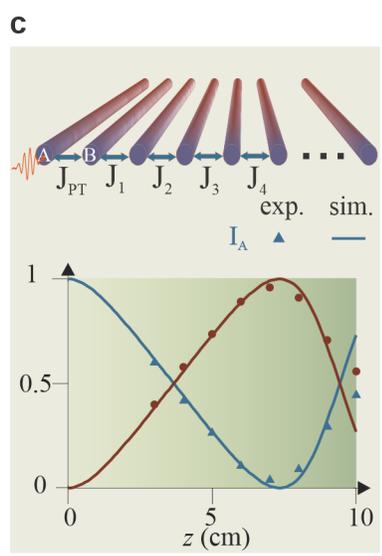

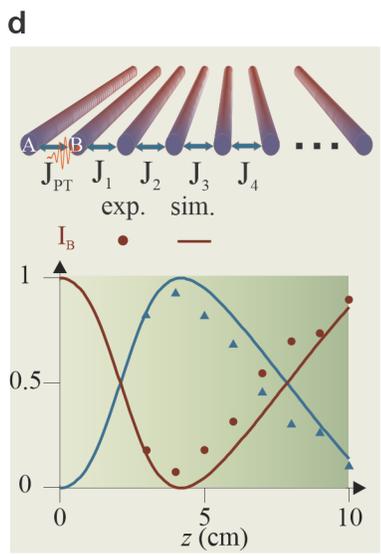

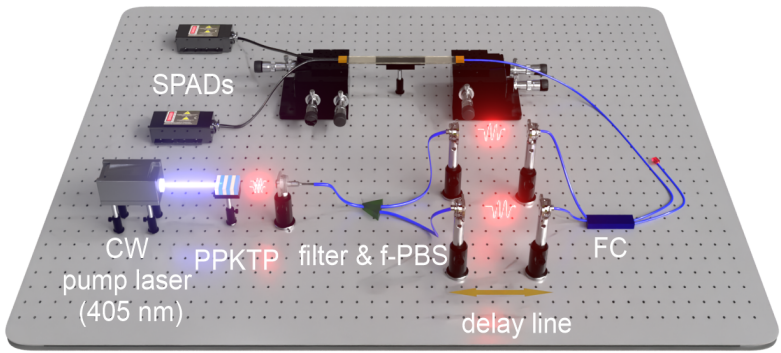

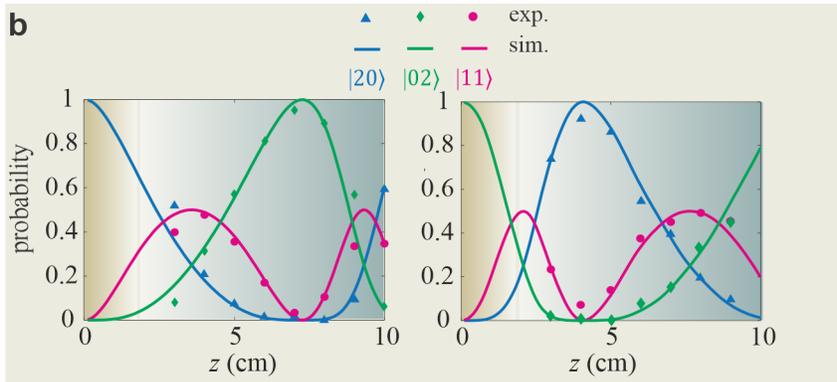

**Correspondence and requests for materials** should be addressed to demetri@usc.edu (D.N.C) and khajavik@usc.edu (M.K.).

**Data and materials availability:** Data are available at Dryad online repositories, at DOI: 10.5061/dryad.2ngf1vhzw.

**Data availability**
Source data are available for this paper. All other data that support the plots within this paper and other findings of this study are available from the corresponding author upon reasonable request.

**Code availability**
The used numerical codes are based upon MATLAB and COMSOL and are available upon reasonable request to the corresponding authors.

**Inclusion & Ethics**
All authors acknowledge the Global Research Code on the development, implementation and communication of this research. For the purpose of transparency, we have included this statement on inclusion and ethics. This work cites a comprehensive list of research from around the world on related topics.

**Additional information**

Supplementary Information is available for this paper.

Correspondence and requests for materials should be addressed to M.K. or D.N.C.

Reprints and permissions information is available at www.nature.com/reprints.